\documentclass[fleqn,10pt'letterpaper]{wlscirep-dfo}
\usepackage[utf8]{inputenc}
\usepackage[T1]{fontenc}
\usepackage{lineno}
\usepackage{notes2bib}
\usepackage{comment}
\usepackage{graphicx}
\usepackage{caption}

\usepackage[numbers]{natbib}
\setcitestyle{super}
\usepackage[labeled]{multibib}
\newcites{S}{Supplemental References}

\title{Localization and reduction of superconducting quantum coherent circuit losses
}

\author[1*]{M. Virginia P. Alto\'{e}}
\author[2,3*]{Archan Banerjee}
\author[1*]{Cassidy Berk}
\author[3,4,5*]{Ahmed Hajr}

\author[1]{Adam Schwartzberg}
\author[1]{Chengyu Song}
\author[6]{Mohammed Al Ghadeer}
\author[1]{Shaul Aloni}
\author[1]{Michael J. Elowson}
\author[2,3]{John Mark Kreikebaum}
\author[1]{Ed K. Wong}
\author[1]{Sinead Griffin}
\author[6]{Saleem Rao}

\author[1]{Alexander Weber-Bargioni}
\author[1,7]{Andrew M. Minor}
\author[3,4]{David I. Santiago}
\author[1]{Stefano Cabrini}
\author[2,3,4]{Irfan Siddiqi}
\author[1$\dag$]{D. Frank Ogletree}

\affil[1]{Molecular Foundry Division, Lawrence Berkeley National Laboratory, Berkeley, CA 94720, USA}
\affil[2]{Materials Sciences Division, Lawrence Berkeley National Laboratory, Berkeley, CA 94720, USA}
\affil[3]{Quantum Nanoelectronics Laboratory, Department of Physics, University of California, Berkeley, CA 94720, USA}
\affil[4]{Computational Research Division, Lawrence Berkeley National Laboratory, Berkeley, CA 94720, USA}
\affil[5]{Graduate Group in Applied Science \& Technology, University of California at Berkeley, Berkeley, CA 94720, USA}
\affil[6]{Department of Physics, King Fahd University of Petroleum and Minerals, Dhahran, Kingdom of Saudi Arabia}
\affil[7]{Department of Materials Science and Engineering, University of California, Berkeley, CA 94720, USA}
\smallskip

\affil[*]{these authors contributed equally to this work}
\affil[$\dag$]{corresponding author: D. Frank Ogletree (dfogletree@lbl.gov)}

\begin{abstract}
    
\end{abstract}
\begin{document}

\flushbottom
\maketitle
\section*{}
\textbf{
Quantum sensing and computation can be realized with superconducting microwave circuits. Qubits are engineered quantum systems of capacitors and inductors with non-linear Josephson junctions. They operate in the single-excitation quantum regime, photons of $27 \mu$eV at 6.5 GHz.  Quantum coherence is fundamentally limited by materials defects, in particular atomic-scale parasitic two-level systems (TLS) in amorphous dielectrics at circuit interfaces.\cite{Muller2019-pn} The electric fields driving oscillating charges in quantum circuits resonantly couple to TLS, producing phase noise and dissipation. We use coplanar niobium-on-silicon superconducting resonators to probe decoherence in quantum circuits. By selectively modifying interface dielectrics, we show that most TLS losses come from the silicon surface oxide, and most non-TLS losses are distributed throughout  the niobium surface oxide. Through post-fabrication interface modification we reduced TLS losses by 85\% and non-TLS losses by 72\%, obtaining record single-photon resonator quality factors above 5 million and approaching a regime where non-TLS losses are dominant.
}

\section*{Background}
TLS are known to be an important decoherence mechanism in quantum circuits, for Josephson junctions as well as for capacitors, inductors, and resonators.\cite{Martinis2005-kp, Gao2008-yi, Wenner2011-zg, Pappas2011-zw} Coplanar waveguide (CPW) resonators are a central component in many superconducting qubit architectures, and also provide a flexible stand-alone vehicle for characterizing microwave frequency materials losses in the single quantum excitation regime.\cite{Muller2019-pn,McRae2020-bh,Burnett2016-qt} 
Resonator geometries\cite{Gao2008-yi, Woods2019-tb} and materials\cite{Wisbey2010-oa, Earnest2018-hm, McRae2020-hx} have been modified to understand and reduce TLS losses associated with the principle resonator material interfaces:  metal-air (MA), metal-substrate (MS), and substrate-air (SA). Elegant spectroscopic experiments using qubits or resonators as quantum sensors have moved beyond the TLS “standard tunneling model”\cite{Phillips1987-jc, Muller2019-pn} to gain insights into TLS-TLS interactions\cite{Lisenfeld2015-yc}, induced fluctuations,\cite{Faoro2012-nw} and noise.\cite{Burnett2014-gx,Burnett2016-qt} Experiments with applied mechanical\cite{Grabovskij2012-of,Lisenfeld2019-ym} or electrical\cite{Sarabi2016-xm,Bilmes2020-ek} strain fields have directly measured TLS electrical and strain dipoles, densities and coupling energies.\cite{Bilmes2020-ek,Sarabi2016-xm,Muller2019-pn}
The relative contributions of different interfaces to TLS and other losses and the atomic-scale structure of TLS remain open questions. Through post-fabrication interface modification of our niobium-on-silicon CPW resonators by selective etching, we have been able to directly associate different loss processes with the Si and Nb surface oxides, and improve the functionality of superconducting circuits by suppressing the majority of these losses.

\section*{Results}
We characterized ten resonator test chips cut from a single 150 mm wafer, each with ten  $\lambda/4$ resonators. X-ray photoemission spectra (XPS) were acquired after etching for variable intervals, followed by microwave reflectometry and scanning transmission electron microscopy (STEM). 

XPS is non-destructive and relatively fast, giving averaged interface information over 400 $\mu m$ regions. XPS can “see through” the surface oxides into the bulk Nb and Si, quantifying the oxide thicknesses and allowing the different Nb and Si oxidation states to be resolved from their chemical shifts.\cite{Halbritter1987-ih,Kuznetsov2009-vs, Oh2001-mj} TEM, on the other hand, is destructive and relatively slow. Cross-section lamella more than 10 microns long and 30 to 50 nm thick were aligned with the crystal lattice of the Si substrate and perpendicular to the CPW axis, typically crossing from the Nb waveguide center conductor, across the exposed Si of the insulating gap, to the Nb ground plane (Fig. S1). High-resolution imaging and TEM nanobeam diffraction revealed the Nb resonator structure. The elemental compositions and chemical states of the three resonator interfaces were mapped using nm-resolution energy-dispersive x-ray fluorescence (STEM-EDS) and angstrom-resolution electron energy loss spectroscopy (STEM-EELS). Accurate determination of film thicknesses by XPS requires morphological models, which are often inferred, however here we have direct knowledge of interface chemistry and morphology from TEM. 

Figure 1 shows the morphology of the standard resonator MA, SA and MS interfaces in panels linked to a cross-section overview and a resonator schematic. The thin slice was lifted out after focused ion beam machining, and shows the edge of the insulating gap where Nb has been etched away. The Si substrate has been over etched by $\sim$ 75 nm relative to the original wafer surface. The sample was aligned with the (011) plane of the Si(100) substrate so TEM images could be acquired on the Si[011] zone axis. The atomic structures and compositions of the three interfaces are analyzed.

The MA interface is shown in detail in the yellow box in Figure 1. The left image is a characteristic region showing the top of several columnar Nb grains, distinguished by TEM contrast.\cite{Kim2013-lm,Premkumar2020-ag} The grain structure creates a surface roughness of a few nm covered with a conformal coating of $\sim$ 5 nm of amorphous niobium oxide. The next panel shows a STEM-EELS Nb/(Nb+O) concentration map, acquired with a 0.1 nm diameter probe. Core-level EELS spectroscopy spatial resolution is limited by the size of the probe. The elemental composition was determined from the Nb $M_4$ and O $K$ edges. The map reveals the NbO$_x$ composition gradient, showing the transition from metallic Nb through an increasingly oxidized phases to Nb$_2$O$_5$ at the air interface. Colors bottom-to-top represent decreasing Nb:Nb+O ratios, corresponding to the Nb$^{\text{0,II,IV,V}}$ oxidation states. EELS spectra (not shown) are consistent with the observed phases.\cite{Bach2006-jw, Tao2011-ra}

The right panel shows the Nb$_{3d}$ XPS peak region. The chemical shifts associated with the different oxidation states\cite{Kuznetsov2009-vs} are large enough to allow the individual contributions to be resolved. The metallic Nb signal is attenuated by the oxide layers, and the ratio of oxidized to metallic Nb peak areas gives an oxide thickness of 4.5$\pm$0.5 nm averaged over $\sim$ 400 $\mu$m. XPS complements the atomic resolution TEM thickness measurements of selected cross-sections. Previous Nb XPS resonator studies proposed a model with a niobium pentoxide surface layer and  sub-oxides near the metal.\cite{Halbritter1987-ih, Premkumar2020-ag} This model is confirmed by our EELS data. A high-resolution SEM image (blue outine) with $\sim$ 25 by 100 nm elongated features reflects the polycrystalline Nb grain structure, consistent with TEM and with scanning probe images.\cite{Premkumar2020-ag}

The SA interface properties are shown in the orange box. There is a sharp transition between the crystalline substrate and the $\sim$ 3 nm SiO$_x$ amorphous oxide. Si$_{2p}$ XPS of the oxide shows a mixture of fully oxygen-coordinated Si$^{4+}$ similar to SiO$_2$ along with sub-oxides with mixed Si-Si and Si-O-Si bonding.\cite{Oh2001-mj}

The MS interface, shown in the green box, is strikingly different from  models commonly described in the literature, which assume an amorphous dielectric interface layer a few nm thick.\cite{Gao2008-yi, Wenner2011-zg, Woods2019-tb, McRae2020-bh} There is no evidence for an amorphous layer in the high-resolution TEM images. Chemical analysis of the interface by STEM-EDS (Fig. S2,S3) and EELS finds only Si and Nb at the interface, with carbon and oxygen below detection limits. Clean Al/Si resonator interfaces have also been reported after substrate processing.\cite{Earnest2018-hm} TEM selected-area diffraction shows an epitaxial relation between the polycrystalline Nb film and the Si(100) substrate. The diffraction data was acquired from a 200 nm diameter region centered on the interface that includes both the Si substrate and a few Nb grains. The incident TEM beam was aligned along the Si [011] “zone axis” of the single-crystal fcc Si lamella. TEM diffraction projects the crystal structures in this plane. In all cases the bcc Nb(011) surface is parallel to the Si(100) wafer surface, but different rotational alignments are observed for different Nb crystals. The yellow lattice in the diffraction pattern is from the Si(011) substrate. The red lattice, projecting the Nb(001) plane, shows the predominant metal orientation, but the blue lattice, projecting Nb(111), is also common. Epitaxial growth requires a clean interface – an amorphous interfacial later would prevent epitaxial growth since Nb nucleation would no longer be sensitive to the substrate crystallography.

Given our clean resonator MS interface, we focused on mitigating process oxides on the MA and SA interfaces through post-fabrication buffered oxide etching (BOE).  We found a 8 pm/s etch rate for niobium oxide, hundreds of times lower than the 1.8 nm/s rate for  silica.\cite{General_Chemical_Electronic_Chemicals_Group2000-ik} A 30 s etch in BOE should completely dissolve the few nm of SA silicon oxide, while removing only a monolayer or so of the thicker MA niobium oxide layer. Extended BOE etching for 600 to 1200 seconds significantly reduced the niobium surface oxide (Fig S4).

The impact of post-fabrication etching was quantified by microwave reflectometry of the frequency-multiplexed test-chip resonators. The observed total quality factor $Q_{T} = f_{0}/\Delta f = \delta_T^{-1}$, where $f_{0}$ and $\Delta f$ are the measured resonant frequency and linewidth at half-maximum and $\delta_T$ is the total loss tangent, including energy lost to the 50 $\Omega$ input feed line ($\delta_{ext} = 1/Q_{ext}$) and internal dissipation ($\delta_{int} = 1/Q_{int}$). Fits to resonator transmission are consistent with $Q_{ext} \sim 0.7 \times 10^{6}$, in good agreement with electromagnetic simulations.  In Figure 2 we compare low and high power reflectometry data for unetched and etched resonators at 100 mK through  resonance curves (2a), the $Q_{int}$ photon number $\langle n \rangle$ dependence (2b), and the $\delta_{int}$ temperature dependence (2c). For unetched resonators we found a median $Q_{int}$ of $0.94 \times 10^6$ at $\langle n \rangle \sim 1$ and $3.2 \times 10^6$ at $\langle n \rangle \sim 10^7$. $Q_{int}$ increased with etch time, reaching a test-chip single-photon median of $5.3 \times 10^6$ and $6.0 \times 10^6$ for the best resonator. In Fig 2d, we plot the low- and high-power losses for each resonator studied along with the NbO$_x$ thickness measured by XPS.

\section*{Discussion}
Although the TLS standard tunneling model is not sufficient to explain resonator noise and power-dependent TLS  saturation, for the current manuscript we will follow a common practice in the literature,\cite{Martinis2005-kp, Gao2008-yi, McRae2020-bh, Woods2019-tb} and decompose the measured resonator internal dissipation $\delta_{int}$ into a power/temperature dependent "TLS" component $\delta_{TLS}$ and a constant hi-power component $\delta_0$ (Fig. 2b). Unsaturated TLS loss channels may contribute to $\delta_0$.

We exploit the $\sim$200:1 BOE etch selectivity between SiO$_x$ and NbO$_x$ to experimentally assign loss contributions to the SA and MA interfaces. 
Standard resonator losses (Fig 3a) are dominated by $\delta_{TLS}$. A short BOE etch dissolves the original SA SiO$_2$ process oxide with minimal impact on the MA Nb$_2$O$_5$. SA interface etching reduces $\delta_{TLS}$ from 71\% to 21\% of the standard resonator reference value. The SA oxide partly regrows after etching and before resonator reflectometry measurements. This regrown ambient oxide is thinner and chemically different from the original process oxide formed in dry conditions, as shown in Fig 3b. Stoichiometric SiO$_2$ is in the 4+ oxidation state, and this component dominated the process oxide, but was suppressed by etching. The regrown SA oxide was $\sim$ 1 nm thick, compared to 2 to 3 nm before etching, and was primarily in the 3+ oxidation state.  The  reduction in $\delta_{TLS}$ suggests that hydroxyl groups, expected with ambient oxidation, may not be a strong contributor to TLS losses in resonator thin film dielectrics. Si$^{4+}$ O-Si-O bonding motifs could be important in thin films, even though OH increases the TLS density in bulk silica.\cite{Phillips1987-jc} It is possible that etching the outermost surface of the MA process oxide also contributes to TLS loss reduction even though XPS does not show significant changes (Fig. 3b).
 
Extended BOE etching reduces the MA NbO$_x$ process oxide. Both $\delta_{TLS}$ and $\delta_o$ decrease monotonically with NbO$_x$ thickness (Fig. 3c). The points at the far right are from the unetched sample, and the sharp initial drop in $\delta_{TLS}$ was due to SA etching. Fig. 2d shows results for individual resonators. With extended MA etching $\delta_{TLS}$ dropped from 21\% to 10\% and $\delta_o$ from 24\% to 8\% relative to the standard resonator. SA and MA etching accounted for at least 80 \% of the single-photon losses for our resonators. SA is the major contributor to $\delta_{TLS}$, while MA is the main source of $\delta_{o}$. 

Long etch times dissolve much of the MA oxide, but etch rate slows as the Nb:O ratio increases. Oxide regrowth on a sputter-cleaned Nb film was $\sim$ 1.3 nm after 90 minutes in air, the approximate time required for mounting a test chip (Figs. S5,S6). This is consistent with the 2$\pm0.5$ nm Nb oxide layer observed after long etch times and cryogenic testing. XPS and EELS data do not show significant changes near the Nb metal/oxide interface (Fig. 3 b,d).

The power and temperature dependence $\delta_{int}$ for low-loss resonators deviate from the predictions TLS standard tunneling model (Fig. S10). The slow saturation with power for the remaining $\delta_{TLS}$ is consistent with energy diffusion resulting from TLS-TLS interactions.\cite{Faoro2012-nw,Burnett2016-qt} The monotonic decrease in $\delta_{TLS}$ and $\delta_0$ with MA oxide thickness shows that the nanoscopic defects responsible for loss are distributed throughout the bulk niobium oxide, and not primarily associated with either the oxide-vacuum or oxide-metal interfaces. Magnetic scattering by niobium sub-oxide defects is one possible source of distributed MA power-independent loss.\cite{Harrelson_undated-os}

\section*{Conclusion}
We have robustly achieved state-of-the-art Q values in niobium co-planar waveguide resonators by removing surface process  oxides in devices with pristine metal-substrate interfaces. Selective etching showed that 
TLS losses  arise primarily from  exposed dielectric process oxides on our resonators, at least 70\% from SiO$_x$ and 15\% from NbO$_x$. Power independent losses are also concentrated in these oxides, 55\% from NbO$_x$ and 17\% from SiO$_x$. As we mitigate surface interface losses, we approach a regime where TLS losses no longer dominate, which could allow losses in superconducting films such a non-equilibrium quasiparticles to be investigated in more detail. Surface passivation after oxide removal, or changes in fabrication to avoid surface oxidation, could lead to general improvements in quantum circuit performance. 

\section*{Methods}

\textbf{Resonator fabrication.} 
Each 10 x 10 mm test chip had ten superconducting niobium  
frequency-multiplexed $\lambda /4$ resonators spanning 6.25 to 7 GHz.\bibnote{Fabricated in the Quantum Nanoelectronics Lab, Department of Physics, University of California, Berkeley}
Gap and width of the CPW waveguide were 10 and 20 $\mu$m respectively.
Resonators were weakly coupled to a 
50 $\Omega$ characteristic impedance feedline with an external quality factor of $\sim$ 700,000 (simulated by Ansys HFSS software and consistent with measurement). Samples were fabricated on double-side polished high-resistivity (> 8000 $\Omega$-cm) intrinsic silicon <100> wafers. Wafers were cleaned in piranha solution (a mixture of sulfuric acid and hydrogen peroxide at  $120^\circ$C) for 10 minutes followed by one minute wet etching in hydrofluoric acid (10:1 solution of 49\% HF) to remove the surface contaminants and native oxides. (Warning: Piranha solutions and HF are hazardous material. Procedures must be carried out by trained workers with proper personal protective equipment and in a fume hood.) Wafers were inserted into an ultra-high vacuum load-locked sputter deposition system immediately after cleaning (process chamber base pressure < $5 \times 10^{-8}$ Torr). Nb films ($\sim$ 180 nm) were deposited by magnetron sputtering at room temperature (deposition pressure 1.5 mTorr, time 10 minutes). A 100 kV Raith  Electron  Beam  Pattern  Generator (EBPG 5150) and MicroChem MMA EL-13 copolymer resist defined circuits. After patterning, wafers were developed in a 3:1 mixture of IPA:MIBK solution at room temperature, then etched in an inductively coupled reactive ion etcher (BCl$_3$/Cl$_2$ process). After etching, residual resist was removed with Mircoposit Remover 1165 at  $80^\circ$C, then wafers were re-coated with MMA EL-13 resist for dicing and storage. 

\textbf{Sample storage, resist stripping and transport.} The protective resist was stripped from test chips by spraying with acetone for $\sim$ 30 seconds, then soaked in a beaker partially filled with Microposit Remover 1165 in an  $80^\circ$C water bath for $\sim$ 12 hours. Chips were then sonicated in deionized water, then acetone, and finally IPA, all at  $50^\circ$C, before and ashing in an oxygen plasma for 1 min. 
Samples were placed in a custom high vacuum transport vessel (stainless steel, KF flanges, dry turbo pump) for transfer between the Molecular Foundry LBL (XPS, TEM/STEM and SEM, cross sectioning, BOE etching) and the Siddiqi labs UCB Physics (chip fabrication, resist stripping,  BOE etching, microwave characterization). No evidence was found of oxide growth or contamination of resist-protected chips during storage.

\textbf{BOE post-fabrication etching.} We confirmed that BOE will etch niobium oxide through XPS experiments on resonators and on sputtered Nb films oxidized by plasma ashing (see supplemental information for addional details). The etch rate was hundreds of times lower than for silicon oxide, $\sim$ 6 pm/s for 5:1 BOE, and 0.7 pm/s for 10:1 BOE, compared to 1.8 and 0.9 nm/s for silica.\cite{General_Chemical_Electronic_Chemicals_Group2000-ik} BOE contains HF, a hazardous material. Etching must be carried out by trained workers with proper personal protective equipment and in a fume hood. Etching at UCB, main text: Four 100 ml cleaned HDPE beakers were used, one containing 15 ml J T Baker 5:1 CMOS grade BOE solution. Chips were handled with teflon tweezers and hand-agitated in BOE for the specified interval, removed from acid and allowed to drip dry, then rinsed for 60 s in each of 3 consecutive rinse beakers filled with DI water, and finally blown dry with nitrogen. Additional Etching at LBL in 10:1 BOE (SI): A PTFE dish was filled with 100 ml of Sigma Aldrich 10:1 BOE and three PP beakers with 100 ml of house DI water. Chips were handled with PTFE tweezers and laid flat in BOE for the specified interval, followed by rinsing for 60 s in three consecutive beakers. In some cases a stir bar was used. In some cases chips had a final rinse in acetone then IPA. The chips were then blown dry with nitrogen.

\textbf{TEM cross-section sample preparation. } TEM cross sections were prepared in a FEI Helios G4 UX dual beam Focused Ion Beam (FIB) equipped with an EasyLift rotatable needle at the National Center for Electron Microscopy (NCEM) facility at the Foundry. Successive layers of electron-beam induced carbon, electron-beam induced platinum and ion-beam induced platinum were deposited at the surface of devices for sample protection before lift-out and thinning. Additional TEM cross sections were prepared by Outermost Technology using similar instruments and procedures.

\textbf{TEM imaging and spectroscopy. } High resolution TEM images and EDS elemental chemical analysis were acquired at 200 kV with a JEOL 2100-F Field Emission Scanning Transmission Electron Microscope (STEM) equipped with an Oxford high solid-angle Silicon Drift Detector (SDD) X-Ray Energy Dispersive Spectrometer (EDS). A 1 nm diameter electron beam probe was used for EDS spectral imaging. Electron Energy Loss Spectroscopy (EELS) was performed at 300 kV in the TEAM I double-aberration-corrected (scanning) transmission electron microscope (STEM/TEM) equipped with a high-resolution Continuum Gatan Imaging Filter (GIF) spectrometer and a 4k x 4k Gatan K3 direct electron detector. EELS spectral imaging maps were acquired using an electron beam probe 0.1 nm in diameter with 1 eV energy resolution (elastic peak).

\textbf{XPS.} X-Ray Photoelectron spectra were acquired with a Thermo-Fisher K-Alpha$^+$ XPS equipped with a focused, monochromatic Al X-ray source, a collection lens with a 30$^\circ$ half-angle acceptance, and hemispherical analyzer with a multi-channel detector at normal incidence to the sample. The X-ray spot size was 200 or 400 $\mu$m depending on the data set. A dual monoatomic/gas cluster Argon ion source was used for sputter cleaning and depth profiling. Peak fitting used "skewvoigt"  for metallic Nb and "pseudovoigt" for other peaks using python software and the Lmfit module.\cite{Newville_undated-at}
A Shirley inelastic background was used (a Tougaard background gave equivalent results). Nb$_{3d}$ 0,2,4 and 5 oxidation state energies were 202.3, 204.1, 206.3  and 207.5 eV respectively. Si$_{2p}$ 0, 2, 3 and 4 oxidation states energies were 99,4, 101.8, 102.7 and 103.5 eV respectively. The Nb:Nb$_2$O$_5$ sensitivity ratio was determined from clean metal and thick oxide samples. The SiO$_x$ effective attenuation length (EAL) was 2.84 nm.\cite{Jablonski2020-or}  The niobium oxide attenuation lengths were calculated with SESSA software\cite{Smekal2005-ub} and rescaled to fit TEM cross section data, resulting in a Nb$_2$O$_5$ EAL of 1.7 nm. 

\textbf{SEM imaging.} Top view SEM images were acquired in a ZEISS Gemini ULTRA-55 Field Emission Scanning Electron Microscope. High magnification images were collected at 2 kV through a high efficiency In-lens Secondary Electrons detector. Low magnification, large field of view images, used 5 kV and a conventional Everhart Thornley Secondary Electron Detector. 

\textbf{Resonator Cryo Characterization.} 
Tests were carried out in a HPD Rainer 103 ADR Cryostat with a base temperature of 65 mK. Most measurements were performed at 100 mK. Measurement and shielding details were similar to Kreikebaum\cite{Kreikebaum2016-qw} except that 
samples in this work were measured in reflection using a microwave circulator. (Fig. S9)
Resonator $Q_{i}$ was fit to $1/Q_i=1/Q-\cos{\phi}/Q_e$ which is obtained from the scattering parameter $S_{11}=A\ e^{i\tau f}\left(1-\frac{2Q/Q_e}{1+2jQ\delta f/f_r}e^{i\phi}\right)$ measured with a network analyzer. See SI and Fig. S10 for fitting details.
Acquisition of a $\langle n \rangle \sim 1$ reflection spectrum for a single resonator required about 8 minutes for good signal to noise.

\bibliography{Paperpile}

\section*{Acknowledgements}
This work was funded by the U.S. Department of Energy, Office of Science, Office of Basic Energy Sciences, Materials Sciences and Engineering Division under Contract No. DE-AC02-05-CH11231 "High-Coherence Multilayer Superconducting Structures for Large Scale Qubit Integration and Photonic Transduction program (QIS-LBNL)" (resonator fabrication, BOE etching, cryogenic characterization, TEM sample prep). Work at the Molecular Foundry was supported by the Office of Science, Office of Basic Energy Sciences, of the U.S. Department of Energy under Contract No. DE-AC02-05CH11231. (SEM, TEM and XPS characterization and data analysis). We acknowledge John Turner for assistance in TEM cross-section preparation, Peter Ercius and Jim Ciston for discussions on EELS data interpretation, and Marie-Paule Delplancke-Ogletree for discussions on inorganic chemistry and metal oxide etching. 

\section*{Author contributions statement}
DFO, IS, AB and MVPA planned the experiments. AB and JMK fabricated the resonators. AB and EW etched resonators. CB and MVPA acquired XPS data. MVPA, CS and SA prepared cross-sections and acquired TEM data. MVPA acquired SEM and optical images. AH and MAG performed microwave reflectometry. EWK fabricated vacuum assemblies.  AH, AB, CB, MVPA, CS and DFO analyzed data. DFO, IS and AS wrote the manuscript and prepared figures. All authors discussed the results and reviewed the manuscript.

\section*{Competing interests} The authors declare no competing financial interests.

\section*{Figures \& Tables}
\begin{figure}[ht]
\centering
\includegraphics[width=\linewidth]{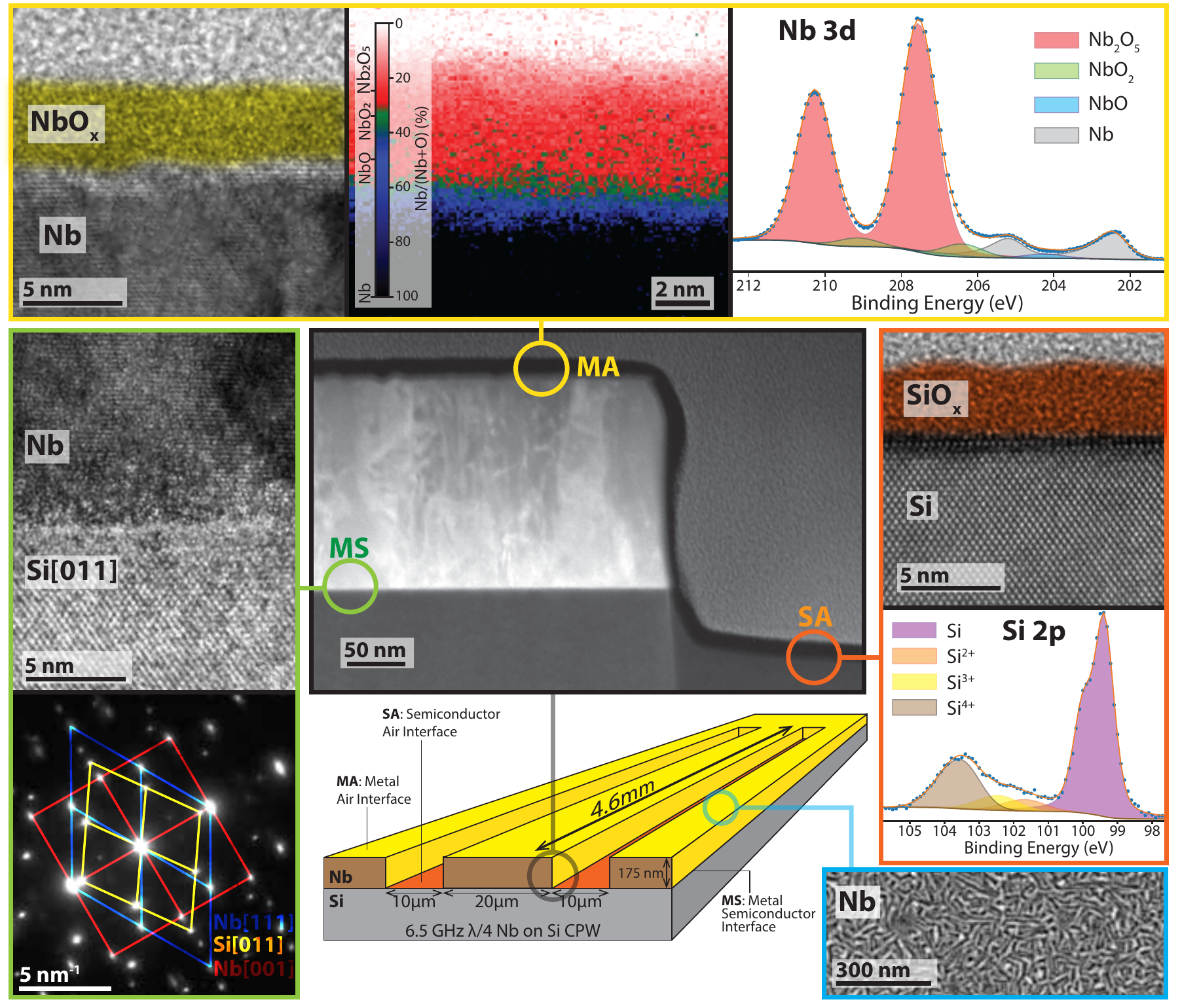}
\caption{Materials analysis of standard process (not etched) resonator interfaces. The Nb on Si CPW resonator schematic (bottom center), is linked to (center) a STEM dark field image of a lift-out cross section showing the MA, SA and MS interfaces. MA detail (top, yellow outline) with from left, HR TEM showing the MA surface roughness due to the Nb grain structure and the $\sim$ 4.5 nm NbO$_x$
amorphous oxide (yellow tint). Elemental STEM EELS map of the Nb/(Nb+O) ratio. Colors bottom to top represent decreasing Nb concentration relative to oxygen.
Nb$_{3d}$ XPS spectrum, with the metallic Nb signal attenuated by the surface oxides. Nanometer resolution SEM (bottom right, blue outline) of the resonator surface with elongated features reflecting the polycrystalline Nb grain structure. SA detail (right, orange box) with a TEM closeup showing $\sim$ 3 nm amorphous SiO$_x$ oxide (orange tint) and the Si$_{2p}$ XPS spectrum with the elemental Si signal attenuated by the surface oxides. MS detail (left, green outline) TEM of the clean MS Nb/Si interface, which is free of carbon and oxygen in EELS and EDS. Epitaxial Nb growth is confirmed by TEM nanobeam diffraction from the interface, which projects crystal structures into the cross-section plane. The yellow lines indicate diffraction spots from the fcc Si(011) substrate plane. Two different epitaxial bcc Nb grains are projected in red (100) and blue (111). Indications of epitaxy are also seen in the TEM image, with Nb atomic rows approximately parallel to the Si substrate.
}
\end{figure}

\begin{figure}[ht]
\centering
\includegraphics[width=\linewidth]{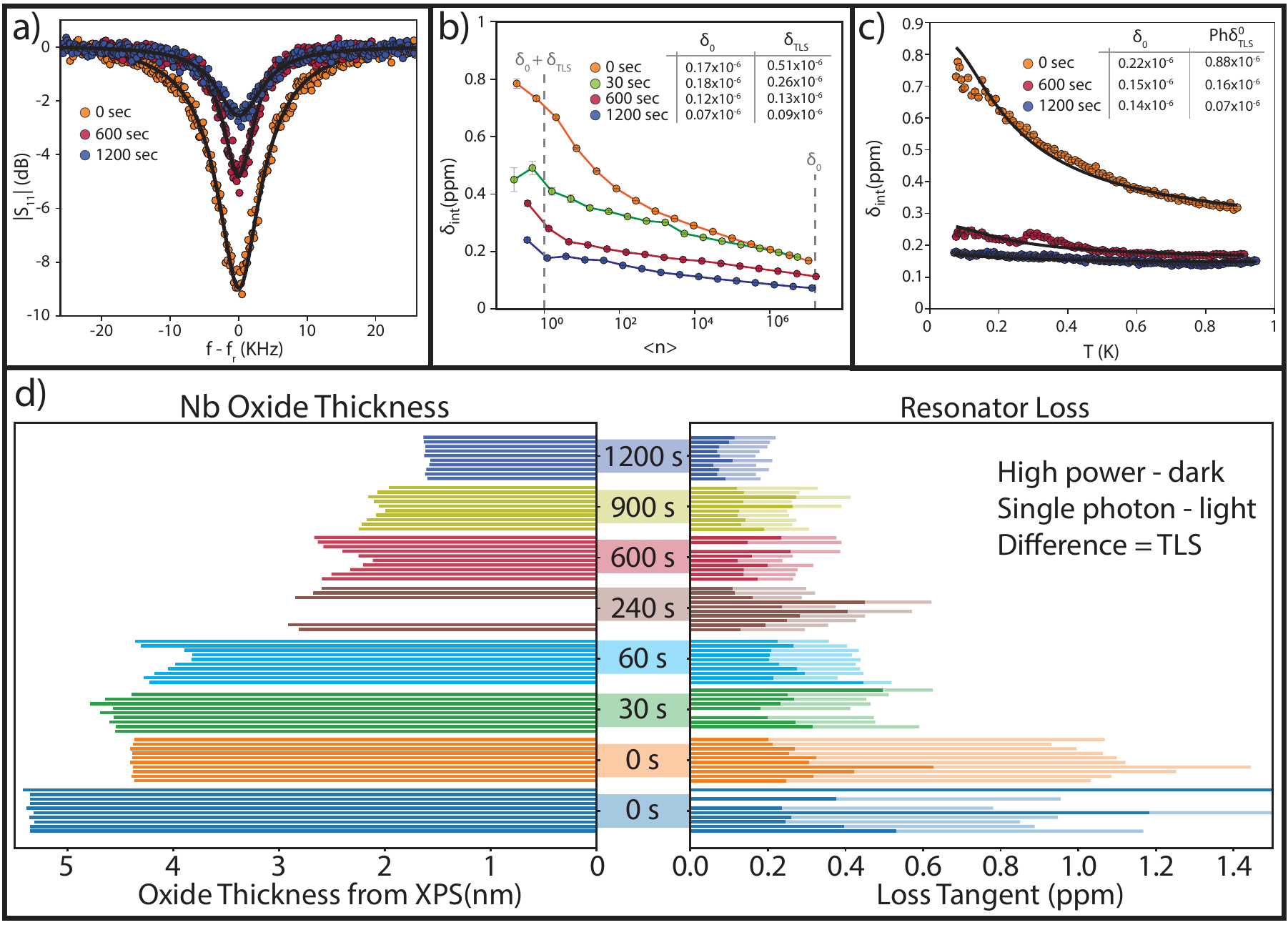}
\caption{(a-c) Comparing the best-performing resonators from unetched (yellow), 600 s (red) and 1200 s (blue) etched chips.
(a) Single-photon $S_{11}$ reflectometry. The decreased peak depth at resonance shows reduced loss.
(b) Resonator loss $\delta_{int}$ as a function of photon number
$\langle n \rangle$ decomposed into the high-power loss $\delta_o$ and single-photon loss $\delta_o + \delta_{TLS}$. Etched resonators show reduced losses at all power levels. 
(c) Temperature dependent $\delta_{int}$ at $\langle n \rangle \approx 1$ fit to the TLS standard tunneling model 
(see methods and SI for details).
(d) Grouped histograms showing resonator loss (right) and Nb oxide thickness from XPS (left) for all resonators. The light-colored $\delta_{int}$
bars show single-photon loss and the darker bars loss for $\langle n \rangle \approx 10^7$ (See also Fig. S11).
}
\end{figure}

\begin{figure}[ht]
\centering
\includegraphics[width=\linewidth]{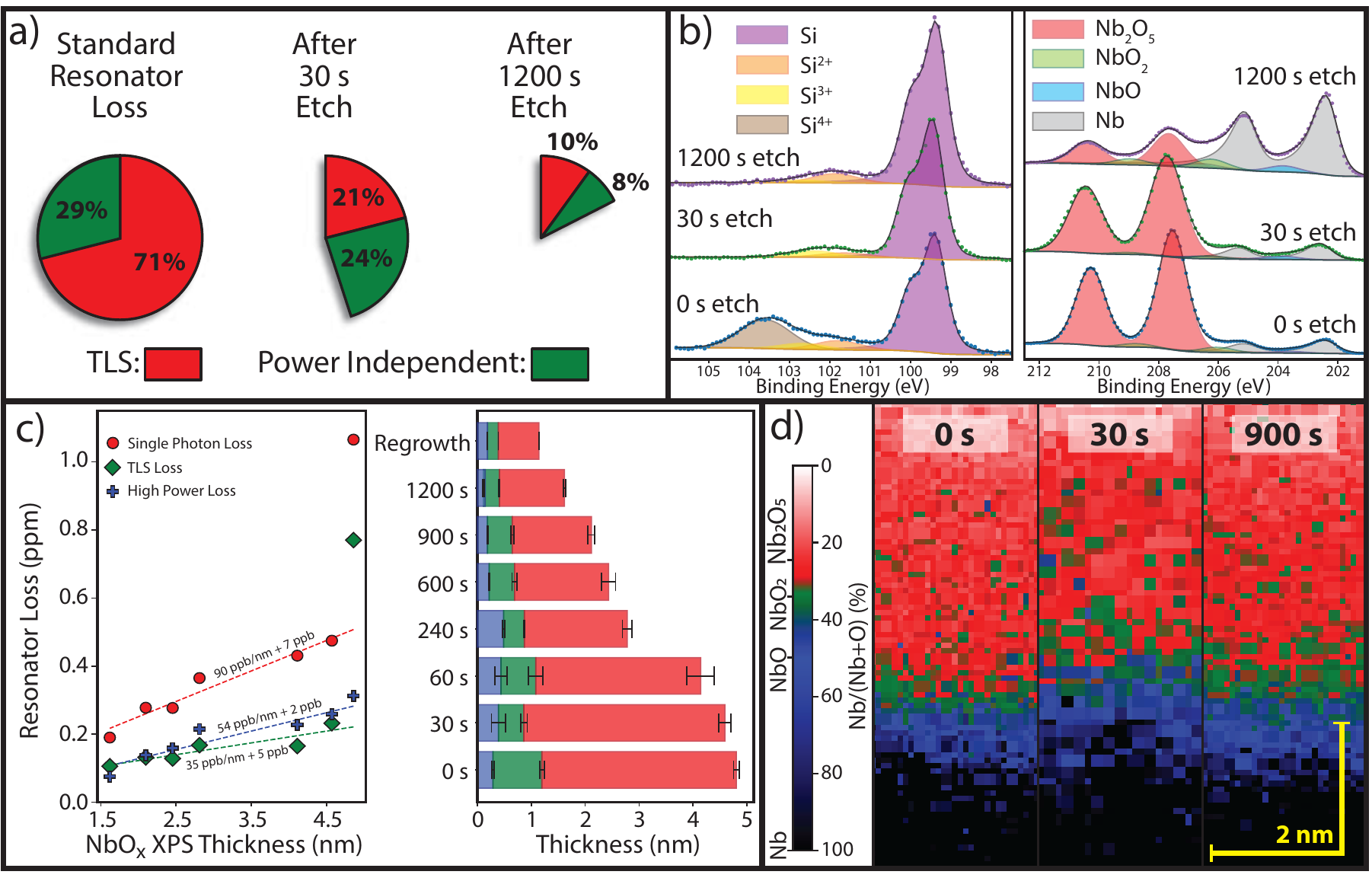}
\caption{Etch effects on resonators.
(a) Pie plots comparing loss distributions before and after etching (areas and percentages are relative to the median standard resonator). TLS and high power loss $\delta_o$ are defined in fig 2b. The standard resonator is dominated by $\delta_{TLS}$. A 30 s BOE etch transforms the silicon SA process oxide and reduces median loss by  55\%. A 1200 s BOE etch also decreases the MA niobium process oxide thickness and reduces loss by an additional 27\%. Residual or regrown oxides may contribute to the remaining 18\% of resonator loss.
(b) XPS data comparing standard, 30 s and 1200 s BOE etched SA (Si$_{2p}$) and MA (Nb$_{3d}$) interfaces. The chemical composition of the regrown SA oxide has changed, and the Si$^{4+}$ component (stoichiometric SiO$_2$) is gone. The MA oxide thickness is significantly reduced after a 1200 s etch, but only slightly modified by the 30 s etch.
(c) Plot of median resonator losses vs NbO$_x$ thickness from XPS with linear fits to the etched data points. The rightmost points represent the unetched resonators. The large drop in $\delta_{TLS}$ with any amount of etching is evident. Remaining losses are distributed throughout the MA oxide. Bar graph showing etching effects on niobium oxide and the thicknesses of the Nb$_{2}$O$_{5}$, NbO$_{2}$ and NbO layers (corrected for photoelectron attenuation). (See also Fig. S8)
(d) Niobium oxide Nb/(Nb+O) EELS maps as a function of etch time showing the spatial distribution of oxide phases. The morphology of the niobium oxide-metal interface does not change significantly even with extended etching.
}
\end{figure}

% To be included, or not, at end of main text

%\clearpage
\renewcommand{\figurename}{Figure S} 
\setcounter{figure}{0}

\begin{center}
{\Large
Supplemental information for

\textbf{Localization and reduction of superconducting quantum coherent circuit losses}
}
\end{center}

\section{Supplemental TEM Data}
STEM analysis was performed on cross-section lamella prepared by focused ion beam machining and extracted from resonator chips (Fig. S-1) both at the Nb ground plane and along the CPW resonator waveguide center conductor. An average of 3 thin sections were obtained from each 10-resonator test chip studied in this work to explore interface structure and variations with etching time. Resonators numbered R1 to R10 on chip.

TEM EDS maps are shown for standard (Fig. S-2) and etched (Fig. S-3) sample lamella. BOE etching effects on the MA interface (Fig. S-4) are shown vs etch time. 

\begin{figure}[h]
\centering
\includegraphics[width=\linewidth]{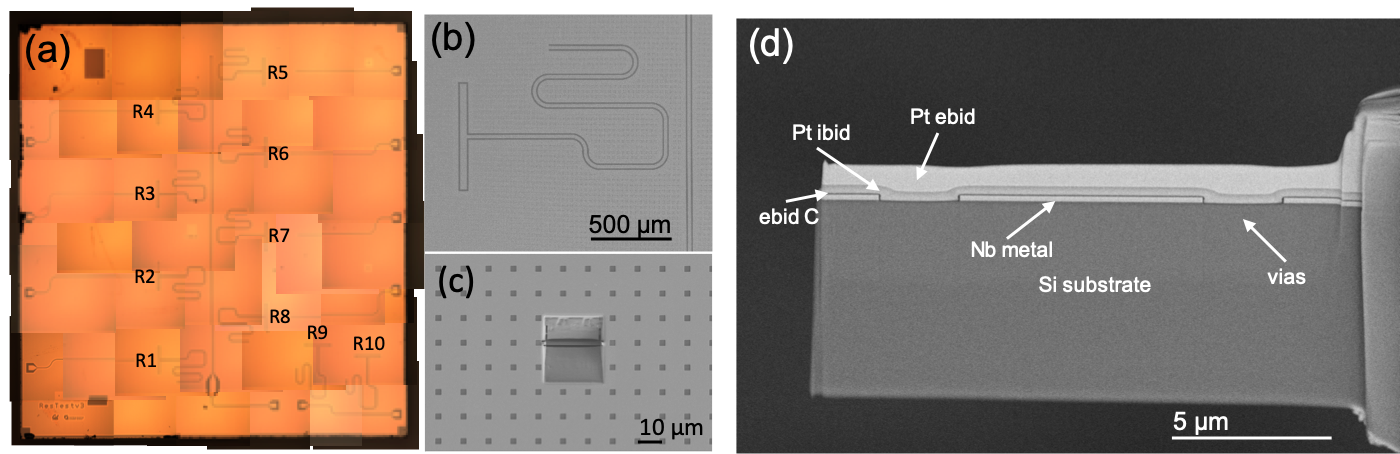}
\caption{TEM sample preparation (a) optical montage of our 10 CPW resonator test chip. SEM images showing (b) top view of resonator R4 and Nb ground plane and (c) after a thin lamella (d) was extracted. Successive layers of carbon (e-beam induced deposition) followed by platinum (electron-beam then ion-beam induced deposition) are deposited to protect the surface from damage during  sample preparation.
}
\label{fig:crosssec}
\end{figure}

\begin{figure}[h]
\centering
\includegraphics[width=\linewidth]{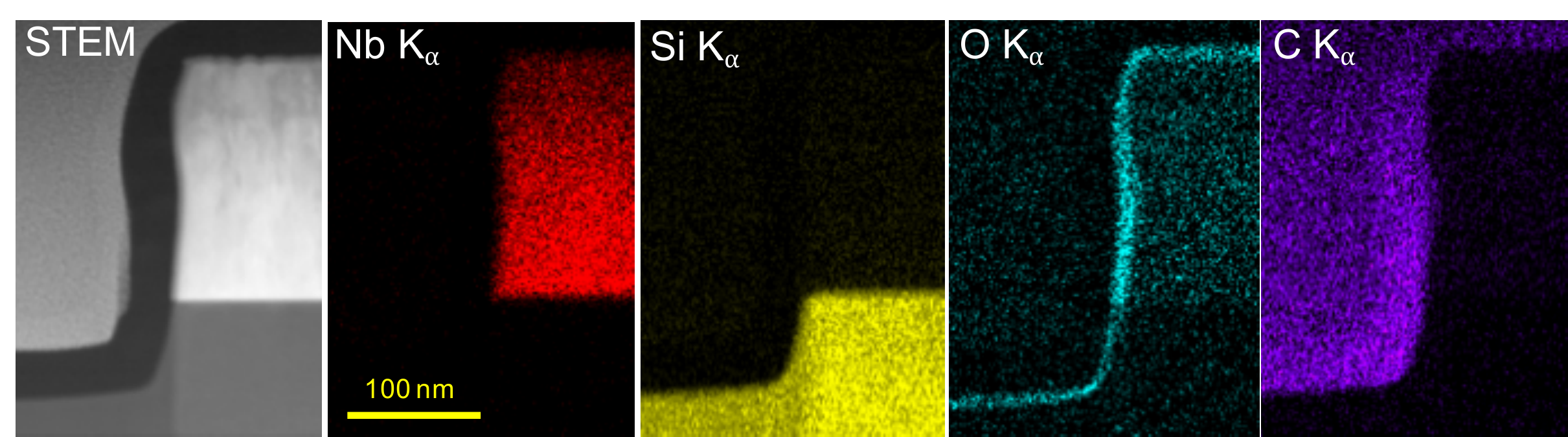}
\caption{Chemical analysis of standard resonator interfaces by TEM Energy Dispersive Spectroscopy (EDS) (x-ray fluorescence). The HAADF (high angle anular dark field) STEM image is shown at left, followed by Nb, Si, O and C elemental concentration images. The MS (metal-substrate) interface is free of C and O. The MA metal-air oxide is thicker than the SA (substrate-air) oxide. During sample preparation monolayer amounts of oxygen and carbon can absorb on the freshly exposed surfaces of the lamella.
}
\label{fig:EDSstandard}
\end{figure}

\begin{figure}[h]
\centering
\includegraphics[width=\linewidth]{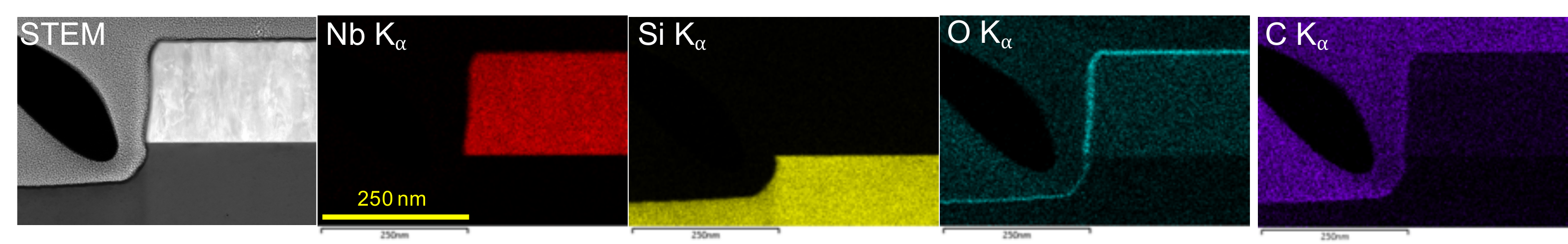}
\caption{ TEM-EDS maps of a chip etched for 10 minutes in 5:1 BOE (Buffered Oxide Etch). Note the reduced oxide thickness at the MA and SA interfaces. The MS interface is again free of C and O.
}
\label{fig:EDS10min}
\end{figure}

\begin{figure}[h]
\centering
\includegraphics[width=\linewidth]{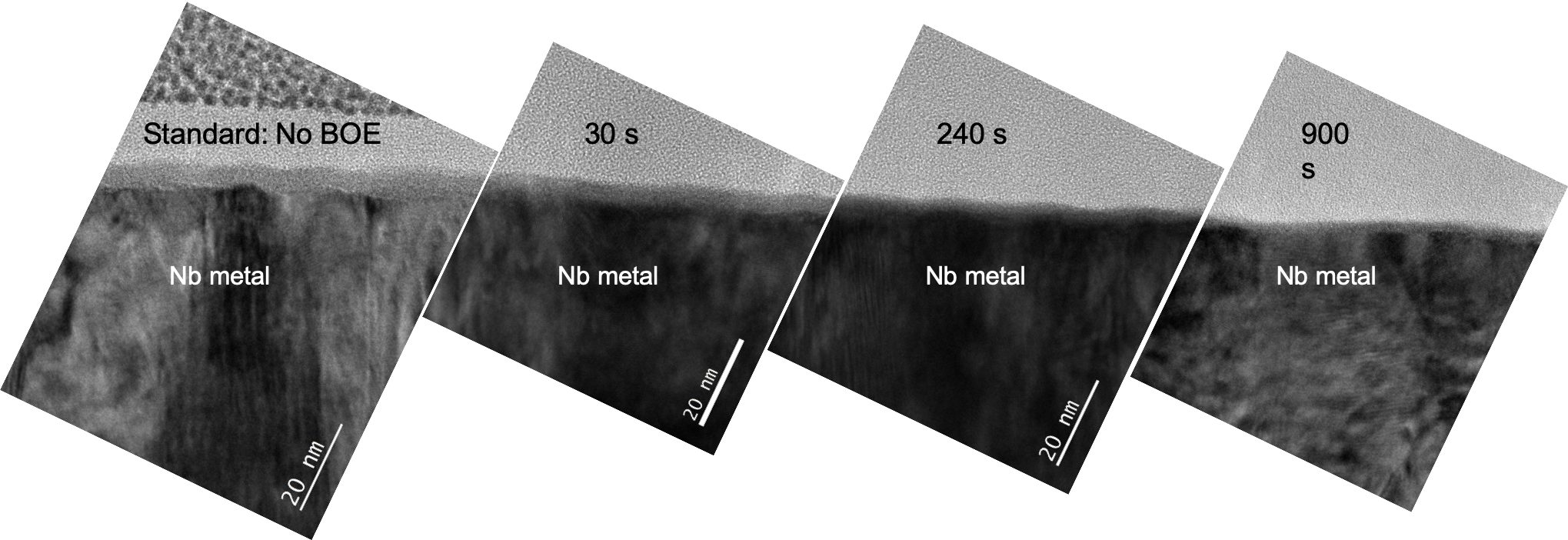}
\caption{ Evolution of the MA interface with BOE etch time. The NbO$x$ layer thickness is steadily reduced with time.
}
\label{fig:TEMma}
\end{figure}

%\clearpage

\section{BOE Etching}
We modified resonators by etching in solutions of “BOE” (buffered oxide etchant), which is used in the semiconductor industry to selectively dissolve silicon oxide layers without attacking the substrate wafers. \cite{General_Chemical_Electronic_Chemicals_Group2000-ik} 

The standard resonator silicon and niobium surface oxides are process oxides, possibly formed during resist processing or reactive ion etching during fabrication. These process oxide layers are stable for many months when protected by the polymer layer used for wafer dicing and storage, or for many days in ambient conditions. The etching experiments presented in the main text used 5:1 BOE and were carried out in the Quantum Nanoelectronics Lab, UC Berkeley Physics department. Some additional etching experiments were performed in the Molecular Foundry Nanofabrication Facility at LBL using 10:1 BOE. See methods in main text for etching details.

\begin{figure}[h]
\centering
\includegraphics[width=4in]{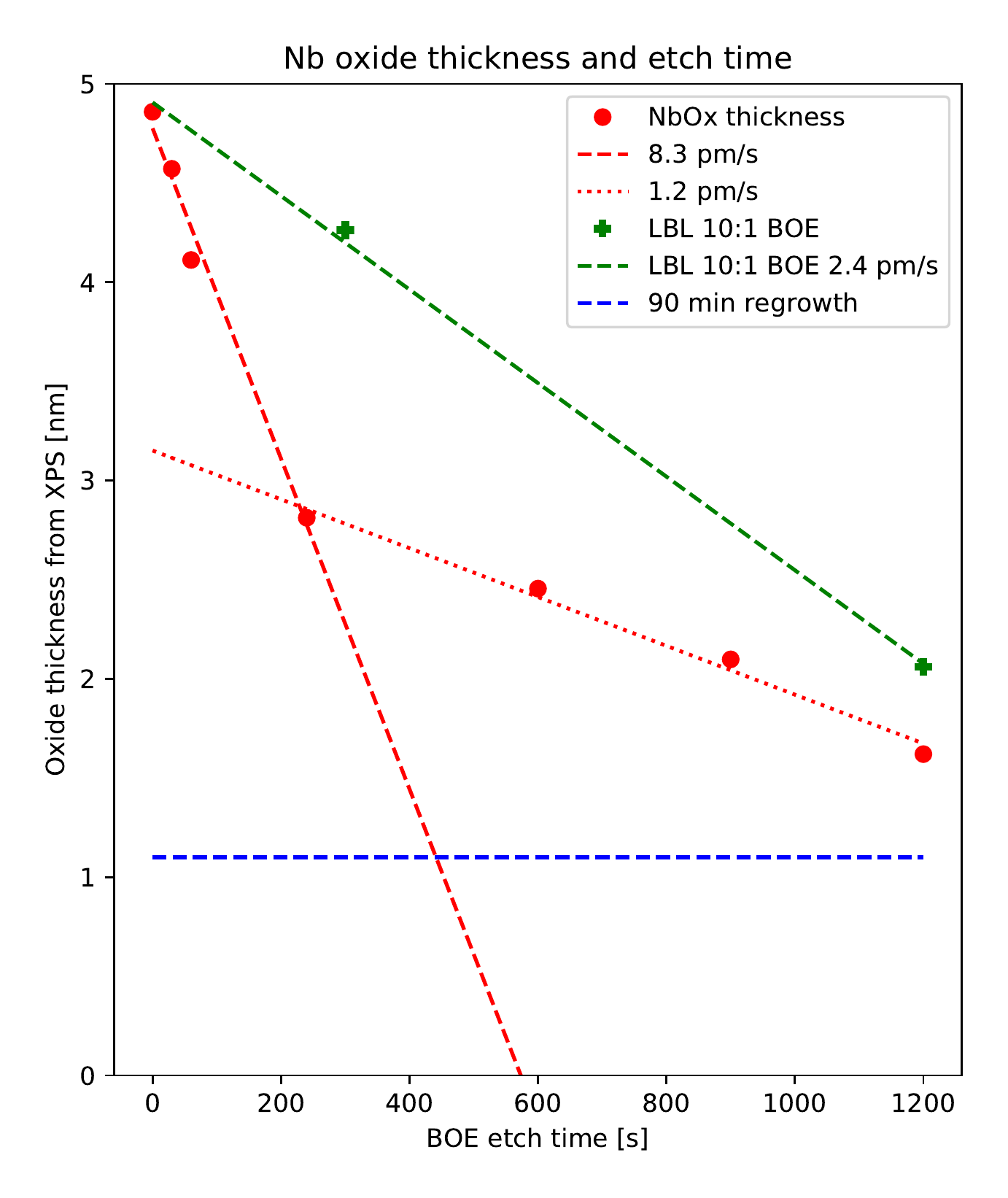}
\caption{ Linear fits to niobium oxide etch rates on resonator samples for 5:1 and 10:1 BOE, determined by XPS. The horizontal dashed blue line indicates the approximate amount of oxide that would grow on clean Nb  during sample mounting for mK resonator characterization. For 5:1 BOE, the etch rate is reduced as the oxide thins and the oxide stoicheometry changes.
}
\label{fig:etchtime}
\end{figure}

Premixed BOE solutions are referred to by ratios like 5:1, a mixture of 5 volumes of 40 wt-\% ammonium fluoride (NH$_4$F) in water to 1 volume of 49 wt-\% hydrofluoric acid (HF) in water.  We used both 5:1 (8.2-8.5 wt\%  HF, 9.9 M NH$_4$F) and 10:1 (4.4-4.7 wt-\% 2.5 M HF, 10.8 M NH$_4$F) BOE solutions. The silicon oxide etch rate has an almost linear dependence on HF concentration. At 21$^\circ$C the etch rates are 1.8 and 0.9 nm/s and increase by $\sim$ 7\%/K.

Ammonium fluoride is more than a buffer to stabilize the HF pH, it plays an active role in SiO$_2$ etch chemistry.  Kikuyama \emph{et al.}\citeS{Kikyuama1991-la} have studied the process in detail. The main active species are HF$_2^-$ and H$^+$. Only about 10\% of the HF dissociates in solution, and only about 10\% of the fluoride ions form the active HF$_2^-$ species in unbuffered solutions. Ammonium fluoride provides F$^-$ to complex with neutral HF and increase the SiO$_2$ etch rate. NH$_4$F alone does not etch SiO$_x$, some H$^+$ from HF is required. 

Mixtures of hydrofluoric, nitric and other acids are used to etch $\sim 100 \mu$m damaged layers from superconducting Nb cavity resonators after fabrication from niobium sheets.\citeS{Ciovati2011-mn, Aspart2004-bi}  Nitric acid oxidizes metallic Nb to surface Nb$_2$O$_5$, then HF reacts with the oxide to form water and soluble NbF$_5$.

We confirmed that BOE will etch niobium oxide. The measured etch rate is hundreds of times lower than for silicon oxide, 8.3 pm/s for 5:1 BOE, and 2.4 pm/s for 10:1 BOE, compared to 1.8 and 0.9 nm/s for silica (Fig. S-5). This large SiO$_2$:Nb$_2$O$_5$ etch selectivity, greater than 200:1 for 5:1 BOE and nearly 400:1 for 10:1 BOE reflects differences in Nb and Si etch chemistry, and allows us to identify the resonator loss contributions from the different interfaces.

The Nb to O ratio increases deeper in the oxide layer and the dominant Nb oxidation state changes from +5 to +4 to +2 as Nb$_2$O$_5$ goes to NbO$_2$ to NbO. This may cause the change in etch rate. The plot in Fig S-5 shows an etch rate change around 2.7 nm of NbOx, while while EELS and XPS peak fitting show the composition changing around 1.5 nm. The difference may be partly caused by the Nb surface roughness of 1-2 nm.

\section{Oxide Regrowth}
After BOE etching samples are rinsed and dried in air then stored in vacuum until they are wire-bonded, cryo-packaged and mounted in the ADR for microwave reflectometry.
In order to better understand the growth kinematics of Nb oxides in atmospheric conditions, XPS analysis was performed on a Nb resonator after Ar$^+$ ion sputter-cleaning in the XPS vacuum chamber. The system pressure in operation was $\sim 10^{-7}$ mBar. After 20 minutes in high vacuum residual oxygen absorbed on the highly reactive Nb metal. The system then was programmed to move the sample to the airlock and expose it to air for a present interval before pumping down for the next XPS spectrum (Fig. S-6).
After 1 minute of exposure to ambient air both NbO$_2$ and Nb$_2$O$_5$ oxides were detected by XPS. The relative thickness of the different oxides were determined by peak fitting and effective attenuation length calculations using a structural model derived from STEM-EELS mapping. All oxides continued to grow with additional exposures. After an hour and 1.1 nm of oxide growth the abmient oxidation rate dropped significantly. Once the Nb$_2$O$_5$ layer was formed, additional oxidation of samples stored in vacuum was negligible.

\begin{figure}[h]
\centering
\includegraphics[width=\linewidth]{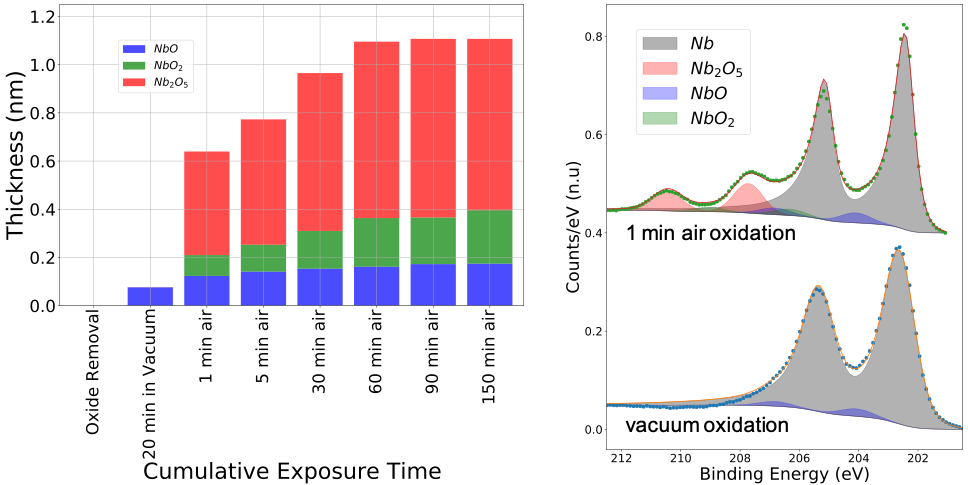}
\caption{ Oxidation of clean Nb in high vacuum and air. After sputter-cleaning in the XPS vacuum, the sample was repeated exposed to air then Nb$_{3d}$ spectra were acquired. A sub-monolayer amount of NbO formed in high vacuum, followed by NbO$_2$ and Nb$_2$O$_5$ with air exposure. After an hour 1.1 nm of niobium oxide had formed, and the oxidation rate decreased significantly. At right are XPS spectra acquired after vacuum oxidation and after the first air exposure.
}
\label{fig:NbRegrowth}
\end{figure}

%\clearpage
\section{BOE 10:1 Etching Data}
In addition to the data for the standard resonators and the chips etched in 5:1 BOE, we also etched one chip in 10:1 BOE for 5 minutes, and another in 10:1 BOE for 20 minutes followed by 1 minute in 5:1. 
Figure S-7 compares the results of XPS and microwave reflectometry for the 10:1 chips to a pair of 5:1 etched chips (data also shown in main text) with similar Nb oxide thicknesses. Except for the difference in etch rates, the loss reductions are consistent with the oxide thickness changes for 5:1 and 10:1 BOE. 

\begin{figure}[h]
\centering
\includegraphics[width=\linewidth]{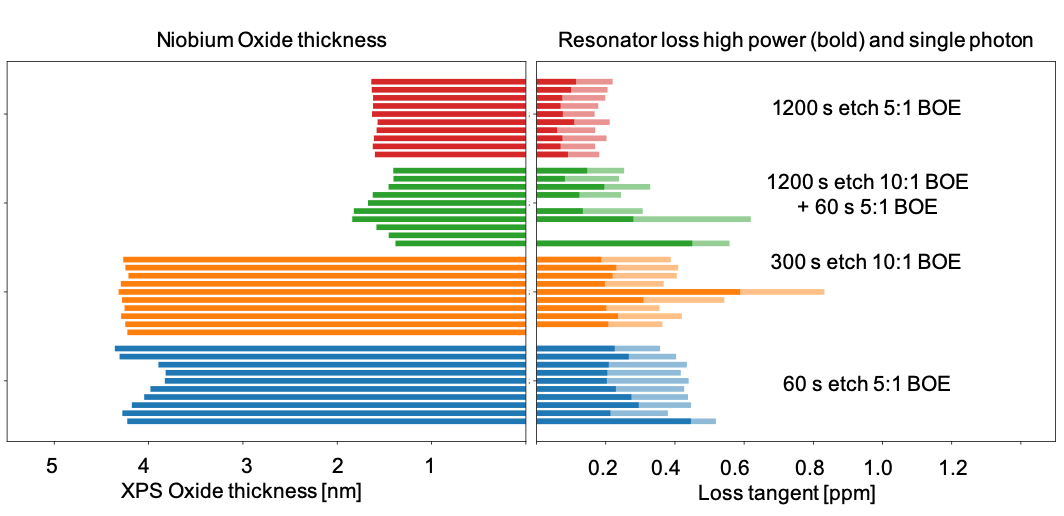}
\caption{ Comparison of chips etched in 10:1 BOE with 5:1 BOE etched chips of similar Nb oxide thickness. One sample was etched in 10:1 BOE only, the other in both 10:1 and 5:1 BOE.
}
\label{fig:10to1}
\end{figure}

\section{Nb Oxide Loss Correlation Plots}
Correlation plots of resonator losses verses niobium oxide thickness determined by XPS for all studied resonators are shown in figure Fig. S-8.

\begin{figure}
\centering
\includegraphics[width=\linewidth]{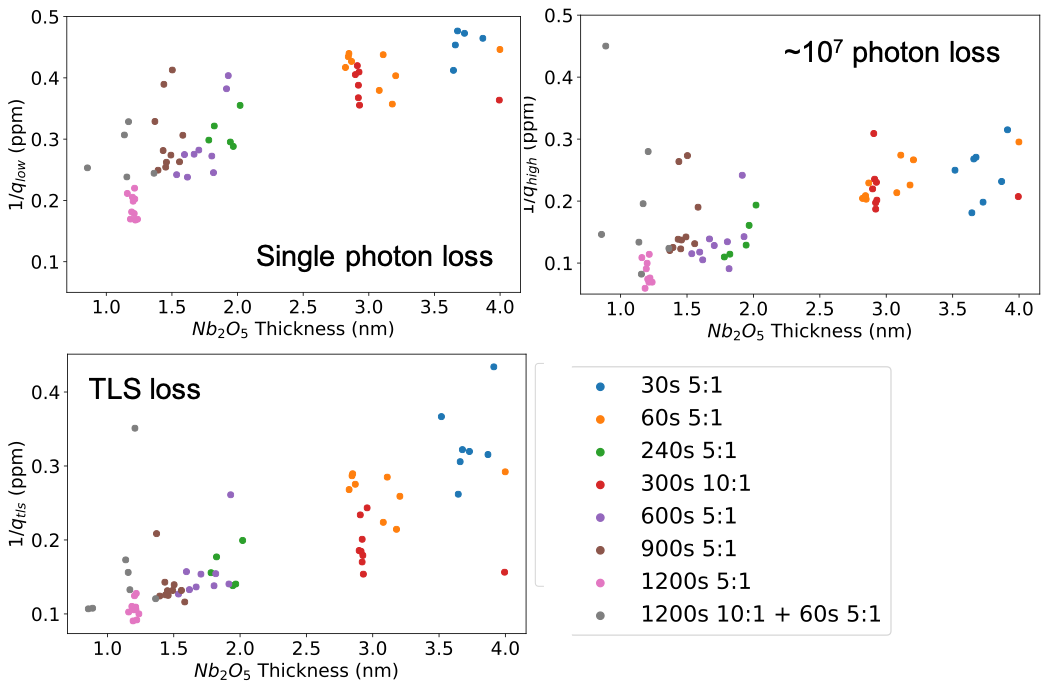}
\caption{ Scatter plots of resonator losses against NbO$_x$ XPS thickness for all etched resonators, color-coded by etch conditions. The unetched resonators are not included, since their losses are more than double that of the etched samples. While some resonators have high losses even with thin MA oxides, thin oxides are necessary for high performance.
}
\end{figure}
\label{fig:correlation}
%\clearpage

\section{Reflectometry Scattering Parameters and Fitting}

The internal quality factor was extracted from the scattering parameters using a vector network analyzer (Rodhe\&Schwartz ZVM). At resonance we observe a sharp attenuation peak which depends on intrinsic resonator loss and external coupling to the feed line.  We designed the chip to achieve an external coupling $Q_e \sim 0.7 \times 10^6$ which was designed to achieve critical coupling at single photon excitation for our standard resonators. The typical model to extract the quality factors from the scattering parameters is a multi-port network (Fig. S-9a) where the coupling is linked to the scattering parameter $S_{31}^\prime$. Port 3 and port 1 refer to the resonator and the feed line respectively and $Q_e \propto 1/|S_{31}^\prime|^2$. The intrinsic quality factor $Q_i$ of the resonator reflects the distributed losses across the $\lambda/4$ resonator. The details of the analysis are provided in \citeS{Gao2008-ke}  where we start with a 3-port network $S_{ij}^\prime$ model of the feed line and the resonator and then calculate the transmission coefficient considering a reduced 2-port network. The scattering parameter $T_{21}$ is given by: 

$$\displaystyle T_{21}=S_{21}^\prime+\frac{S_{31}^\prime\Gamma\ S_{23}^\prime}{1-\Gamma \: S_{33}^\prime}\approx1-\Delta $$

$$\displaystyle \Delta=\frac{Q/Q_{ext}}{1+2jQ \: \delta f/f_r}e^{i\phi} $$

\begin{figure}[h]
\centering
\includegraphics[width=\linewidth]{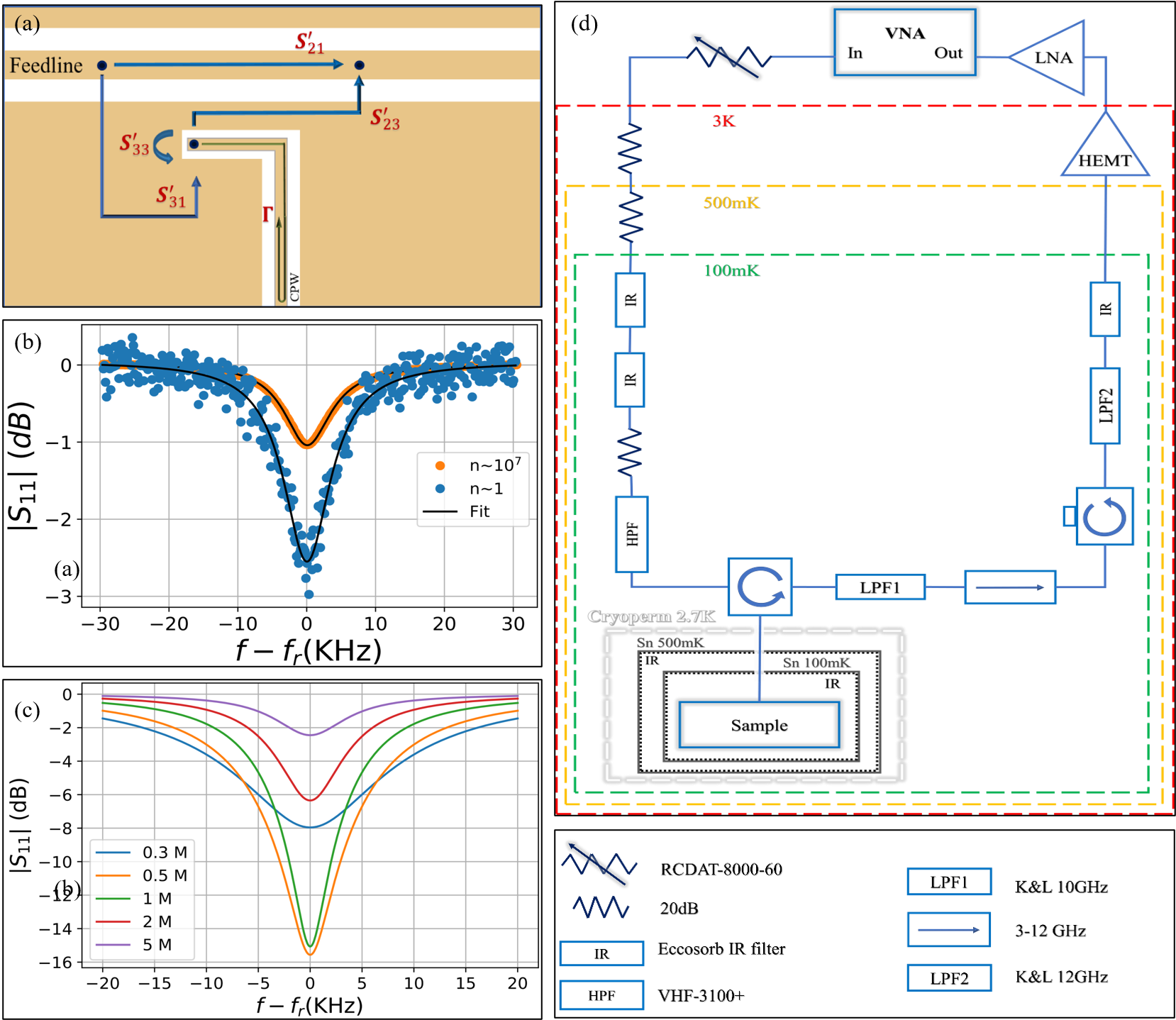}
\caption{	(a) Schematic of the three-port network and scattering parameters. (b) The reflected signal, after normalization for $A$ and $\tau$, at high power and in the single photon region for one of the best standard resonators (R6). There is no significant asymmetry in the background. (c) Plots of $|S_{11}|$ calculated for several values of $Q_{int}$ with $Q_{ext} = 0.7 M$. The power dip is maximized for $Q_{int} \approx Q_{ext}$. The full width at half max increases for $Q_{int} < Q_{ext}$
and approaches a minimum for $Q_{int} > Q_{ext}$. (d) Schematic of the ADR thermal and radiation shielding and microwave components for reflectometry. The resistor symbols show thermalized attenuators, and IR, HPF and LPF are infrared, high-pass and low-pass filters.
}
\label{fig:ADR}
\end{figure}

We measure the reflected signal when the other end of the feed line is open ($Z_L\rightarrow\ \infty,\ \Gamma_L\rightarrow1)$. An intuitive way to understand this configuration is to consider the resonator attenuating the signal twice,  first when it passes by as an incoming signal from the left, and the second time from the right after being reflected at the open end of the feed line. So the reflection coefficient $S_{11}$ in the reduced 1-port measurement is given by
$ \displaystyle S_{11}=T_{21}T_{12}\approx1-2\Delta $
where only the linear term in $\delta f$ is kept and the result is identical to a capacitively coupled resonator at the end of the feed line (1-port model). A more rigorous derivation \citeS{Bockstiegel2016-qq} is to correct for the extended part of the feedline on the network first by introducing the reflection coefficient $\Gamma_f$, and then consider a resonator measured in reflection, the final 
result is given by: 

$$\displaystyle S_{11}\ =\Gamma_f\ \left(1-\frac{2\frac{Q}{Q_{ext}}e^{i\phi}}{1+2jQ \frac{\delta f}{f_r}}\right)
$$
The extended part of the feed line modifies the coupling, so that if $Q_e^\prime$ is the coupling without an extension, we get $Q_e=Q_e^\prime\frac{4\Gamma_f}{\left(1+\Gamma_f\right)^2}$. The factor $e^{i\phi}$ is introduced to take into account the impedance mismatch \citeS{Gao2008-ke, Bockstiegel2016-qq}\cite{McRae2020-bh} and this formula is further modified to include a delay time $\tau$ and a fixed attenuation level at the output of the network analyzer $A$ which includes $\Gamma_f$:

$$\displaystyle S_{11}=A\ e^{i\tau f}\left(1-\frac{2Q/Q_{ext}}{1+2jQ \: \frac{\delta f}{f_r}}e^{i\phi}\right) $$

However, as pointed out in \cite{McRae2020-bh},\citeS{Khalil2012-xg} $Q_i$ should be extracted using $1/Q_i=1/Q_i-\cos{\phi}/Q_e$ which prevents overestimating $Q_i$ in the case of a   strong coupling mismatch. In our case we find negligible asymmetry. This can be seen in Fig S-9b which shows the fits for R6 at high and low power levels, normalized for $A$ and $\tau$, and  determined with the lmfit minimizer package.\cite{Newville_undated-at} Figure 2a in the main text shows $S_{11}$ fits using these formulas. 

The input power can be converted to average photon number using  $\langle n \rangle = \frac{2}{\pi  hf^2}\frac{Q^2}{Q_e}P_{in}$ which can be derived by setting the difference between the input and the reflected power equal to the dissipated power  $P_{in}=P_{in}\left|S_{11}\right|^2+2\pi f{E}/Q_{int}$. The input power was estimated from the network analyzer output, taking into account the attenuation from a RCDAT-8000-60 programmable attenuator and  $\sim 100$ dB of attenuation from the components inside the fridge (measured at room temperature). Fig. S-9d shows a schematic of the set up. More details on the shielding can be found in Kreikebaum. 
\emph{et al.}\cite{Kreikebaum2016-qw} The input line was attenuated by 20 dB and thermalized at each stage and passed through homemade absorptive infrared blocking filters (ecosorb) and a Mini-Circuits VHF-3100+ provided additional filtering. 

The output line included two K\&L 10 and 12 GHz low pass filters, a homemade absorptive infrared blocking filter, 36 dB of isolation from two circulators, and 40+36 dB of gain from a LNC4-8A HEMT at 2.7 K and a MITEQ AFS4-00100800-14-10P-4 amplifier at room temperature.

%\clearpage
\section{Resonator Response Fitting}

The TLS standard tunneling model predicts a temperature-dependent resonant frequency shift
\cite{Phillips1987-jc}
$$\displaystyle \frac{\Delta f_r}{f_r} = 
\frac{p_h\ \delta_{TLS}^0}{\pi}
\left[Re\ \Psi\left(\frac{1}{2}-\frac{h \: f_r}{j2 \pi k_B T}\right)-\log{\left(\frac{h \: f_r}
{k_BT}\right)}\right]$$
where $p_h$ is the participation ratio, the fraction of electric field energy stored in the TLS host material, and $\delta_{TLS}^0 $ is the loss tangent. This response is independent of the saturation of resonant TLS or interactions between TLS. The same model predicts the temperature dependence of TLS loss in the low power limit.
$$\displaystyle \delta_{TLS}(T) = p_h \delta_{TLS}^0 
\tanh \left(\frac{h f_r } {2k_B T} \right)$$
In a real quantum circuit with multiple TLS host materials their contributions are summed.

The TLS standard tunneling model power dependence is: \cite{McRae2020-bh} 
$$ \displaystyle \delta_{TLS}(T,\langle  n \rangle) = 
\frac{\delta_{TLS}(T)}
{ \sqrt{ 1 +
\left( \frac{\langle  n \rangle}{n_{c}}  \right) ^\beta}}$$ 
where $n_c$ is the photon number where saturation effects reduce loss by $\frac{1}{\sqrt2}$.
For uniform electric fields as in parallel plate capacitors or Josephson junctions $\beta \sim 1$. For distributed structures such as a CPW resonator where electric fields vary with distance from the resonator electrodes and along the length of the resonator the effective value of $\beta$ can be smaller, but not less than $\sim$ 0.8 for spatially uniform TLS distributions.\citeS{Khalil2011-cf} Fig. S-10(a) shows an example with $\beta \sim 0.25$, a clear deviation from the standard model, and an indication of TLS energy diffusion due to TLS-TLS interactions. \cite{Faoro2012-nw,Burnett2016-qt} 

Fig. S-10(d) shows observed fluctuations $Q_{int}$ for the 1200 s BOE etched chip over 10 consecutive measurement cycles over 13 hours. The best and worst resonators on the chip are not the same in different cycles. We will continue to investigate the factors contributing to fluctuations.

\begin{figure}
\centering
\includegraphics[width=\linewidth]{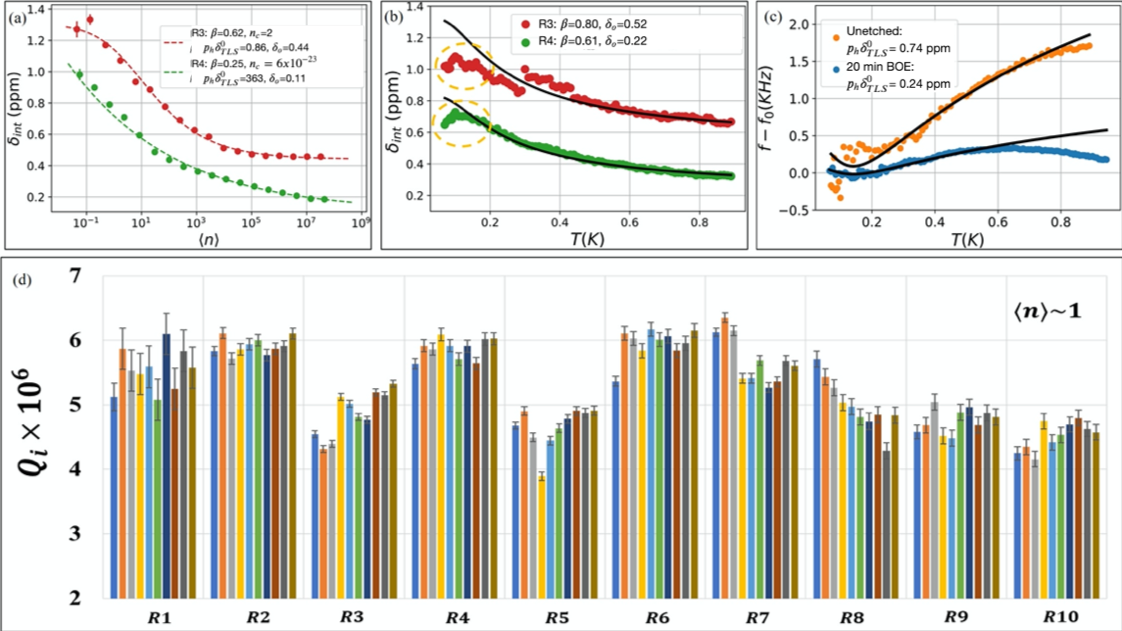}
\caption{ (a) Photon number dependence of $\delta_{int}$ for resonators R3 and R4 from an unetched chip. Some resonators, typically with higher losses such as R3, are in reasonable agreement with the standard tunneling model, while the model clearly fails for others such as R4, where 
$\delta_{int} \propto \langle n \rangle^{-0.12}$. (b) Temperature dependence of $\delta_{int}$ for the same resonators. A deviation from the standard model at low temperature is observed for some resonators both etched and unetched, indicated by the dashed yellow circle. (c) Temperature dependent frequency shifts for the best unetched (yellow) and 1200 s BOE etched (blue) resonator, R6 in both cases. TLS frequency shifts are due primarily to non-resonant TLS interactions, and are not affected by saturation or TLS-TLS interactions.\cite{Phillips1987-jc} BOE reduces the non-resonant TLS contribution three-fold. (d) Repeated measurements of $Q_{int}$ for all resonators on the 1200 s BOE etched chip in sequence. The measurement cycle was repeated 10 times for 100 total measurements, with each $Q_i$  measurement taking 480 s for a total time of 13.5 hours. The blue bars are the first set of measurements, the orange the second, etc. Error bars show fitting uncertainty. 
}
\label{fig:fits}
\end{figure}

\section{Resonator loss table}
The table in Fig. S-11 shows Single Photon,  $> 10^7$ Photon and TLS  median losses for each test chip studied. "TLS loss" is defined as the difference between Low and High power loss. The XPS derived Nb and Si oxide thicknesses are also given. NbO$_x$ thickness decreases monotonically with etch time. SiO$_x$ thickness fluctuates from chip to chip, which we attribute to variations in environmental exposure after etching.

\begin{figure}
\centering
\includegraphics[width=\linewidth]{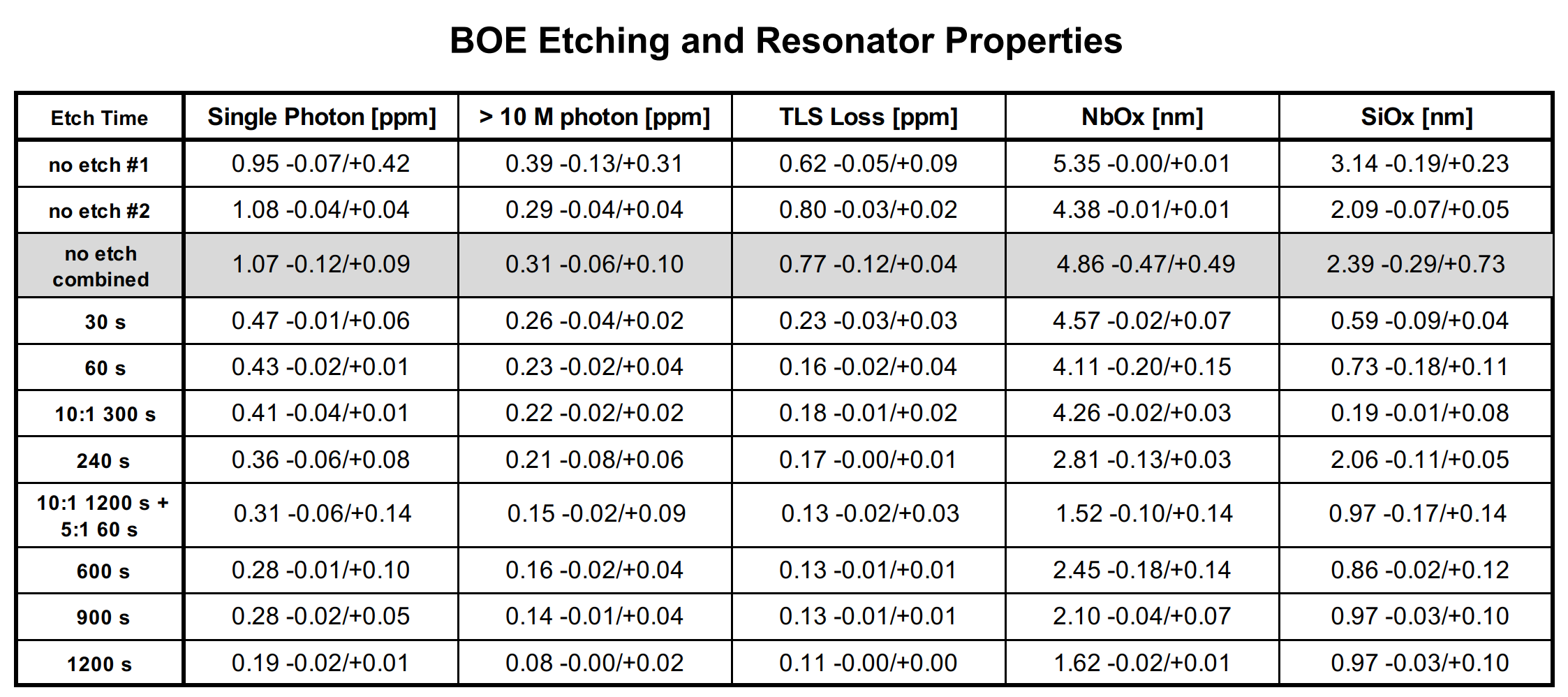}
\caption{ Tabulated median values of etched chips as a function of etching conditions. Columns are single photon, > $10^7$ photon and TLS loss, followed by the niobium and silicon oxide thicknesses determined by XPS. The shaded horizontal row has the combined median properties of the tested standard resonators. The $\pm$ ranges are first and third quartiles. Medians and are quartiles are used instead of mean and standard deviations since some chips had outliers, individual resonators with values far from the median which skew the mean/standard deviation values. Medians and quartiles are less sensitive outlier effects.
}
\label{fig:table}
\end{figure}

%\clearpage
\bibliographystyleS{apsrev}
\bibliographyS{Paperpile}
%\bibliography{Paperpile}

\section*{}
Citations not found here refer to references in the main text.

\end{document}